\documentclass[conference]{IEEEtran}

\usepackage{graphicx}
\usepackage{lipsum}
\usepackage{hyperref}
\usepackage{cite}
\usepackage{pifont}
\usepackage{enumitem}
\usepackage[font=small,skip=0pt]{caption}
\usepackage{adjustbox}
\usepackage{subcaption}

\usepackage[font=small,skip=0pt]{caption}
\IEEEoverridecommandlockouts
\usepackage{atbegshi}
\usepackage{orcidlink}
\AtBeginDocument{\AtBeginShipoutNext{\AtBeginShipoutDiscard}}

\begin{document}

\title{\LARGE Multi-Beam Object-Localization for Millimeter-Wave ISAC-Aided Connected Autonomous Vehicles 
}

\author{\small
Jitendra~Singh$^{\orcidlink{0000-0002-0951-4097}}$, \textit{Graduate Student Member,~IEEE,}
Awadhesh~Gupta$^{\orcidlink{0000-0002-7345-6389}}$, \textit{Graduate Student Member,~IEEE,}
Aditya~K.~Jagannatham$^{\orcidlink{0000-0003-1594-5181}}$, \\ \textit{Senior Member,~IEEE}
and Lajos~Hanzo$^{\orcidlink{0000-0002-2636-5214}}$, \textit{Life Fellow,~IEEE}
}
\thanks{J. Singh A. Gupta and A. K. Jagannatham are with the Department of Electrical Engineering, Indian Institute of Technology Kanpur, Kanpur, UP 208016, India (e-mail: jitend@iitk.ac.in; awadheshg20@iitk.ac.in; adityaj@iitk.ac.in).}
\thanks{L. Hanzo is with the School of Electronics and Computer Science, University of Southampton, Southampton SO17 1BJ, U.K. (e-mail: lh@ecs.soton.ac.uk).}

\maketitle
\begin{abstract}
 Millimeter wave (mmWave) multiple-input multiple-output (MIMO) systems capable of integrated sensing and communication (ISAC) constitute a key technology for connected autonomous vehicles (CAVs). In this context, we propose a multi-beam object-localization (MBOL) model for enhancing the sensing beampattern (SBP) gain of adjacent objects in CAV scenarios. Given the ultra-narrow beams of mmWave MIMO systems, a single pencil beam is unsuitable for closely located objects, which tend to require multiple beams. Hence, we formulate the SBP gain maximization problem, considering also the constraints on the signal-to-interference and noise ratio (SINR) of the communication users (CUs), on the transmit power, and the constant modulus of the phase-shifters in the mmWave hybrid transceiver. To solve this non-convex problem, we propose a penalty-based triple alternating optimization algorithm to design the hybrid beamformer. Finally, simulation results are provided for demonstrating the efficacy of the proposed model.
\end{abstract}

\begin{IEEEkeywords}
Integrated sensing and communication, millimeter wave, connected autonomous vehicles, sensing beampattern gain. 
\end{IEEEkeywords}

\maketitle

\section{\uppercase{INTRODUCTION}}
\IEEEPARstart{C}{onnected} autonomous vehicles(CAVs) have generated significant interest as a key 6G application \cite{rev_3,mm_ISAC_4}, wherein an integrated sensing and communication (ISAC)-enabled base station (BS) provides sensing and communication services to autonomous vehicles in order to avoid collisions. Recent works \cite{cha_est,mm_2,mm_ISAC_5,mm_ISAC_6,mm_ISAC_7} have explored millimeter wave (mmWave) multiple-input multiple-output (MIMO) technology to achieve the dual goals of high sensing beampattern (SBP) gains at the objects and ultra-high data rates at the communication users (CUs) in ISAC systems. 
The authors of \cite{mm_ISAC_5,mm_ISAC_6,mm_ISAC_7} have successfully developed fully-digital beamforming techniques for vehicular applications of ISAC systems. Moreover, due to the mobility of vehicles, it is exceedingly challenging to track them by an ISAC system. For mitigating this challenge, the authors of \cite{mm_ISAC_5} have proposed an extended Kalman
filtering (EKF) framework to estimate and track the kinematic parameters of high-mobility vehicles in wireless networks. As a further advance, the authors of \cite{mm_ISAC_6} designed a novel Bayesian technique for achieving this objective. 
To improve the tracking accuracy of objects and the QoS of the CUs for an arbitrary road geometry, the authors of \cite{mm_ISAC_7} introduced a pioneering curvilinear coordinate system for ISAC-enabled vehicular networks.


The above-mentioned contributions \cite{mm_ISAC_5,mm_ISAC_6,mm_ISAC_7} optimize the beamforming for a point object. However, this assumption only holds true for mmWave MIMO ISAC systems having a sufficiently large distance between the ISAC BS and the object. Yet, in practical CAV scenarios, the distance of the object may be small, hence the object may occupy a wide angular range. Therefore, a single ultra-sharp pencil beam is often inadequate for localizing an object and fails to provide a reliable communication link for the CAVs. To overcome this, Du \textit{et al.,} \cite{mm_ISAC_8} proposed a novel ISAC-based predictive beam tracking technique having alternate wide and narrow beams (ISAC-AB), assuming a wide beamwidth for object-localization and a narrow beamwidth for the CUs. Moreover, the authors of \cite{mm_ISAC_8} have not considered the hybrid beamforming, which is a critical requirement for mmWave ISAC-enabled CAV scenarios.
To this end, Qi \textit{et al.} \cite{mm_ISAC_1} have proposed a two-stage method for optimizing hybrid beamforming in mmWave MIMO ISAC systems, which aims to match the transmit beam with the desired beam pattern of the closely located object by relying on the signal-to-interference and noise power ratios (SINRs) of the CUs as the quality of service (QoS) constraints.
However, the schemes of \cite{mm_ISAC_8,mm_ISAC_1} are capable of tracking a closely located object thanks to its wide beamwidth, albeit this is not a power-efficient solution since increasing the beamwidth inevitably results in reducing the array gain.
\begin{table*}[t!]
    \centering
    \caption{Contrasting our novel contributions to the literature of mmWave MIMO ISAC systems} \label{tab:lit_rev}
    \begin{adjustbox}{width=0.8\linewidth}
\begin{tabular}{|l|c|c|c|c|c|c|c|c|c|c|c|c|c|}
    \hline
 &\cite{rev_3}  &\cite{mm_ISAC_4}  &\cite{cha_est} &\cite{mm_2}  &\cite{mm_ISAC_5}  &\cite{mm_ISAC_6}   &\cite{mm_ISAC_7}   &\cite{mm_ISAC_8}  &\cite{mm_ISAC_1}  &Proposed \\ [0.5ex]
 \hline
ISAC systems   &\checkmark   &\checkmark  &   &    &\checkmark  &\checkmark  &\checkmark  &\checkmark &\checkmark  &\checkmark \\
\hline
mmWave MIMO system   &    &\checkmark   &\checkmark  &\checkmark  &  &\checkmark  &\checkmark  &\checkmark  &\checkmark  &\checkmark  \\
\hline
Hybrid beamforming    &    &\checkmark   &\checkmark  &\checkmark  &  &   &   &   &\checkmark  &\checkmark \\
 \hline
SINR as QoS    &   &  &  &   &   &  &   &   &\checkmark  &\checkmark  \\
 \hline
{\bf MBOL model}  &   &   &   &  &  &   &  &  &  &\checkmark \\
 \hline
{\bf Weighted SBP gain maximization}   &   &   &   &  &   &  &   &  &  &\checkmark\\
 \hline
{\bf Penalty-based triple alternating optimization}   &  &  &  &  &  &   &  &  & &\checkmark \\
 \hline
\end{tabular}
\end{adjustbox}
\end{table*}

In order to address these challenges, we propose a multi-beam object-localization (MBOL) framework for enhancing the SBP gain of closely located objects in an mmWave MIMO ISAC system. These multiple beams form a rough contour of the object, which increases the accuracy of sensing the surrounding vehicles for a CAV. To achieve this goal, we formulate the pertinent optimization problem (OP) for maximizing the weighted SBP gain of the MBOL model, imposing also the practical constraints arising due to the minimum SINR requirements of the CUs, the total transmit power, and the hardware restrictions of the hybrid mmWave MIMO architecture at the ISAC BS. The resultant problem is highly non-convex due to the presence of non-convex constraints and coupled variables. To decouple the variables and handle the non-convex constraints, we introduce an auxiliary variable for the SINR constraint and propose a penalty-based triple alternating algorithm to iteratively optimize the BB and RF precoders, which involves manifold optimization and second-order cone programming (SOCP). 
The simulation results demonstrate that for the given SINR requirements, the SBP gain of the proposed model is higher than that of the benchmark schemes, which renders it well-suited for implementation in practical CAV systems. 
The novel contributions of this work are boldly contrasted to the existing studies in Table \ref{tab:lit_rev} at a glance.

We use the following notation.
The variables $\mathbf{A}$,  $\mathbf{a}$, and $a$ represent a matrix, a vector, and a scalar quantity, respectively;
The $(i,j)$th element, and $i$th element of a matrix $\mathbf{A}$ and a vector $\mathbf{a}$ are denoted by $\mathbf{A}{(i,j)}$ and $\mathbf{a}(i)$, respectively. The Hermitian, transpose and conjugate transpose of a matrix $\mathbf{A}$ are denoted by $\mathbf{A}^H$, $\mathbf{A}^T$ and $\mathbf{A}^*$, respectively; $\left\vert\left\vert \mathbf{A} \right\vert\right\vert_F$, $\left\vert\left\vert \mathbf{a} \right\vert\right\vert_2$ and $\left\vert a \right\vert$ denote the the Frobenius norm, $l_2$ norm and magnitude, respectively.
$\mathbf{A} \odot\mathbf{B}$ is the Hadamard product of $\mathbf{A}$ and $\mathbf{B}$; $\nabla f$ denotes the gradient vector of the function $f$; the real part of a quantity is denoted by $\Re\{\cdot\}$; ${\mathbf I}_M$ denotes an $M \times M$ identity matrix; the symmetric complex Gaussian distribution of mean $a$ and variance $\sigma^2$  is represented as ${\cal CN}(a, \sigma^2)$.
\section{System Model and Problem Formulation}\label{mmWave MU MIMO CR System}
\begin{figure}[t]
\vspace{-1.5cm}
\centering
\includegraphics [width=8cm]{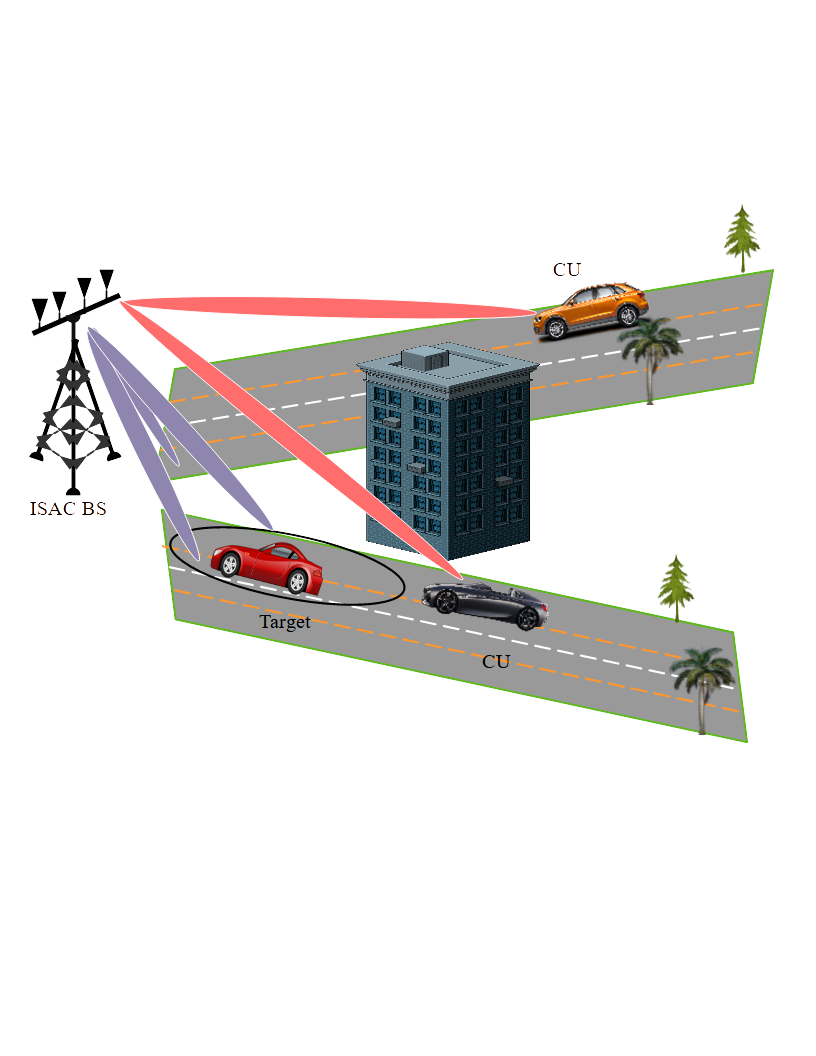}
\vspace{-2.3cm}
\caption{MBOL model for the mmWave ISAC assisted CAVs.}
\label{figure:Fig1}
\end{figure}
As shown in Fig. \ref{figure:Fig1}, we consider a CAV scenario, where an ISAC BS provides services for autonomous vehicles operating in the mmWave frequency band. 
The ISAC BS equipped with $N_\mathrm{t}$ transmit antennas (TAs)/ receive antennas (RAs) transmits $K=L+M$ streams to detect a closely located object with $L$ different beams. Simultaneously, the ISAC BS serves $M$ CUs, each having a single antenna. A fully connected hybrid architecture is exploited at the ISAC BS with only $M_\mathrm{t}\leq N_\mathrm{t}$ RF chains to reduce the cost and power consumption. Notably, for the model considered, a minimum of $M_\mathrm{t}\triangleq K$ RF chains are required at the ISAC BS to form the $L$ different beams required for the objects and the $M$ beams desired for the CUs. 
To this end, we define the transmit signal $\mathbf{x}\in \mathbb{C}^{K \times 1}$ as $\mathbf{x} = [\mathbf{s}^T_1 \hspace{2mm}\mathbf{s}^T_2]^T$,
where $\mathbf{s}_1=[s_1, s_2,\hdots, s_{M}]^T \in \mathbb{C}^{M\times 1}$ is intended for the CUs and $\mathbf{s}_2=[s_{M+1},\hdots, s_{M_\mathrm{t}}]^T\in {\mathbb C}^{L \times 1}$ is employed for object-detection. Furthermore, we assume that both the signals meant for CUs and the object are statistically independent with zero mean, i.e., $ \mathbf{x}$ satisfies the properties $\mathbb{E}\{\mathbf{x}\}=\mathbf{0}$ and $\mathbf{E}\{\mathbf{x}\mathbf{x}^H\}=\mathbf{I}_{K}$.
It is worth noting that the signaling for some CUs may also contribute to the task of object detection.
Following the fully-connected hybrid architecture of \cite{mm_2}, the transmitted signal $\mathbf{x}$ is first precoded by the BB TPC $\mathbf{F}_\mathrm{BB}=[\mathbf{f}_{{\rm BB},1}, \hdots,\mathbf{f}_{{\rm BB},K}]\in {\mathbb C}^{{M_\mathrm{t}} \times K}$, followed by the RF TPC $\mathbf{F}_\mathrm{RF}=[\mathbf{f}_{{\rm RF},1}, \hdots,\mathbf{f}_{{\rm RF},M_\mathrm{t}}]\in {\mathbb C}^{{N_\mathrm{t}} \times {M_\mathrm{t}}}$ prior to transmission. 
\subsection{Radar model}\label{Radar model}
Since the ISAC BS simultaneously transmits downlink signals and receives echo signals, we assume that the receiver is equipped with $N_\mathrm{r}$ receive antennas, which are sufficiently separated from the transmit array for directly suppressing the interference emanating from the transmitter \cite{rev_1}.
Considering the object-tracking mode of the radar system \cite{mm_ISAC_4,mm_ISAC_1,mm_ISAC_8,ISAC_3}, the radar signal received at the ISAC BS can be written as
\begin{equation} 
\mathbf {y}_{rad}=\sum_{l=1}^L\tau_l\mathbf{a}_\mathrm{r}(\theta_l)\mathbf {a}_\mathrm{t}^{T}(\theta _{l})\mathbf{F}_{\rm RF}\mathbf{F}_{\rm BB}\mathbf{x}+\mathbf {n}_{r}, 
\end{equation}
where the first term is the desired target signal and $\mathbf {n}_{r} \in \mathbb{C}^{N_\mathrm{r}\times 1}$ represents the noise signal in the radar sensing environment. Furthermore, $\tau_l$ is the reflection coefficient, $\mathbf{a}_\mathrm{t}(\theta_l)\in \mathbb{C}^{N_\mathrm{t}\times 1}$ and $\mathbf{a}_\mathrm{r}(\theta_l)\in \mathbb{C}^{N_\mathrm{r}\times 1}$ are the transmit and receive array response vectors, which are given by
\begin{equation}\label{eqn:array respo}
\begin{aligned} 
\mathbf{a}_{z}\left (\theta_l \right)=\frac {1}{\sqrt{N_z}}\Bigl [1, \ldots, e^{j \frac {2 \pi }{\lambda } d \left(n \cos \theta_l\right)} ,\ldots, e^{j \frac {2 \pi }{\lambda } d \left ({N_z-1)\cos \theta_l}\right)}\Bigr]^{T},
\end{aligned}
\end{equation}
where $z\in \{\rm r,t\}$. The quantity $\lambda$ denotes the wavelength and $d$ is the antenna spacing which is assumed to be half of the wavelength. 
In order to detect the object, the ISAC BS scans the space to generate multiple beams toward the object. Furthermore, the maximum beamwidth in the mmWave MIMO system does not exceed $\frac{\pi}{2N_\mathrm{t}}$ 
and hence, there is negligible interference among the different beams.
To enhance the sensing performance of the radar, we evaluate the linear combination of the SBP gains, which is given by
\begin{equation}\label{eqn:tx beam pattern}
\begin{aligned}
\Psi(w_l, \mathbf{F}_\mathrm{RF}, \mathbf{F}_\mathrm{BB}) & = 
\sum_{l=1}^{L}w_l\left\vert\mathbf{a}^H_{\mathrm{t}}(\theta_l)\mathbf{F}_\mathrm{RF}\mathbf{F}_\mathrm{BB}\mathbf{F}_\mathrm{BB}^H\mathbf{F}_\mathrm{RF}^H\mathbf{a}_{\mathrm{t}}(\theta_l)\right\vert^2,
\end{aligned}
\end{equation}
where $w_l$ is the weighing factor obeying $\sum_{l=1}^L w_l=1$.
\subsection{Communication model}
Upon assuming full CSI availability at each CU through pilot symbols \cite{cha_est}, the received signal $y_m$ of the $m$th CU is given by
\begin{equation}\label{eqn:rx signal_1}
\begin{aligned}
y_m=&\mathbf{h}^H_m \mathbf{F}_{\rm RF}\mathbf{F}_{\rm BB}\mathbf{x} + n_m, 
\end{aligned}
\end{equation}
where $\mathbf{h}_m$ represents the narrowband block-fading mmWave MISO channel between the ISAC BS and the $m$th CU, which is formulated as
\begin{equation}\label{eqn:channel}
\mathbf{h}^H_{m}= \sum_{i=1}^{N^{\rm p}_m}\alpha_{m,i}\mathbf{a}_{\rm t}^H(\theta_{m,i}), 
\end{equation}
where $N^p_m$ denotes the number of multipath components in $\mathbf{h}_m$. The quantity $\alpha_{m,i}$ is the channel gain of the $i$th multipath, which is an independent random variable having the distribution $\mathcal{CN}(0,\gamma_m^210^{-0.1PL(d_m)}), \forall l=\{1,\hdots, N^p_m\}$, where $\gamma_m=\sqrt{N_\mathrm{t}/N^p_m}$ denotes the normalization factor with $PL(d_m)$ as the pathloss that depends on the distance $d_m$ associated with the corresponding link. The quantity $n_m\sim\mathcal{CN}(0, \sigma^2)$ is the complex circularly symmetric white Gaussian noise. With the aid of (\ref{eqn:rx signal_1}), the SINR of the $m$th CU can be expressed as
\begin{equation}\label{eqn:8}
\mathrm{SINR}_m = \frac{\left\vert\mathbf{h}^H_m \mathbf{F}_{\rm RF}\mathbf{f}_{{\rm BB},m}\right\vert^2}
{\sum_{n=1, n \neq m}^{M_\mathrm{t}}{\left\vert\mathbf{h}_m^H \mathbf{F}_{\rm RF}\mathbf{f}_{{\rm BB},n}\right\vert^2} + \sigma^2 }.
\end{equation}
\subsection{Problem formulation}\label{problem formulation}
Here, we aim for maximizing the weighted SBP gain $\Psi(w_l, \mathbf{F}_\mathrm{RF}, \mathbf{F}_\mathrm{BB})$ of the object that is allocated $L$ radar beams. This is achieved by optimizing the hybrid precoder $\mathbf{F}_\mathrm{RF}$ and $ \mathbf{F}_\mathbf{BB}$. The pertinent OP is given by
\begin{align}
& \max_{\mathbf{F}_\mathrm{RF}, \mathbf{F}_\mathbf{BB}}\Psi\Big(w_l, \mathbf{F}_\mathrm{RF}, \mathbf{F}_\mathrm{BB}\Big) \label{eqn:system optimization} \tag{7a}\\ 
& \text {s.~t.} \quad
\mathrm{SINR}_m \geq \Gamma_m, \forall m, \label{constr:SINR} \tag{7b} \\
&\qquad \left\vert\mathbf{F}_\mathrm{RF}(i,j)\right\vert = 1, \forall i, j, \label{constr:HBF} \tag{7c}\\
&\qquad \|\mathbf{F}_\mathrm{RF}\mathbf{F}_\mathrm{BB}\|_F^2\leq P_\mathrm{t}, \label{constr:TP} \tag{7d}
\end{align}
where $\Gamma_m$ denotes the required SINR as the QoS requirement of the $m$th CU and $P_\mathrm{t}$ is the maximum transmit power at the ISAC BS. Moreover, (\ref{constr:HBF}) is the constant modulus constraint imposed on the elements of the RF TPC due to the hybrid mmWave MIMO architecture. The OP (7) is challenging to solve due to the non-convex nature of the constraints (\ref{constr:SINR}), (\ref{constr:HBF}), and owing to the tightly coupled variables $\mathbf{F}_\mathrm{RF}$ and $\mathbf{F}_\mathrm{BB}$. 
\section{Proposed solution}\label{blind MMSE}
To solve the highly non-convex OP $(8)$, we propose a penalty-based triple alternating optimization algorithm, in which we first decouple the variables by employing the penalized method and thereby optimize the variables using the alternating optimization technique. 

Toward this end, let us introduce the auxiliary variable $\zeta_{m,n}$ to represent $\zeta_{m,n}=\mathbf{h}_m^H\mathbf{F}_\mathrm{RF}\mathbf{f}_{\mathrm{BB},n}, \forall m, n$. As a result, the SINR constraint (\ref{constr:SINR}) can be equivalently written as
\begin{align} 
\frac { \left |{\zeta_{m,m} }\right |^{2}} {\sum_{n \neq m}^{K}\left |{\zeta_{m,n} }\right |^{2}+\sigma_{m}^{2}} \geq \Gamma_{m},\quad \forall m, \label{constr:SINR_1} \tag{8a}\\
\zeta_{m,n}= \mathbf{h}_m^H\mathbf{F}_\mathrm{RF}\mathbf{f}_{\mathrm{BB},n},\quad \forall m, n. \label{constr:SINR_2}\tag{8b}
\end{align}
Observe that while the variables are decoupled in $(9)$, it is still non-convex due the equality constraint (\ref{constr:SINR_2}). To overcome this impediment, we add the equality constraint ($\ref{constr:SINR_2}$) to the objective function as a penalty term.
Furthermore, exploiting the orthogonality of the RF TPC vectors, the transmit power constraint (\ref{constr:TP}) can be written as 
\begin{equation}\label{constr:TP_1}
\|\mathbf{F}_\mathrm{RF}\mathbf{F}_\mathrm{BB}\|_F^2 = N_\mathrm{t}\|{\mathbf{F}_\mathrm{BB}}\|_F^2 \leq P_\mathrm{t}.
\tag{9}
\end{equation}
Note that all the variables $\mathbf{F}_\mathrm{RF}, \mathbf{F}_\mathrm{BB}$ and $\zeta_{m,n}$ are separable and therefore can be optimized alternatively.

Upon adding the equality constraint (\ref{constr:SINR_2}) to the objective function, the resultant OP can be equivalently written as
\begin{align}
&{\min \limits_{{\mathbf{F}_\mathrm{RF}, \mathbf{F}_\mathrm{BB}, \zeta_{m,n}}}} - \sum_{l=1}^L\sum_{m=1}^{M_\mathrm{t}}w_l\left\vert\mathbf{a}^H_\mathrm{t}(\theta_l) \mathbf{F}_\mathrm{RF}\mathbf{f}_{\mathrm{BB},m}\right\vert^2 \notag\\
&\qquad \qquad \qquad + \frac {\alpha}{2} \sum_{n=1}^{M_\mathrm{t}} \sum_{m=1}^{M_\mathrm{t}} \left | \mathbf{h}_{m}^{H} \mathbf{F}_\mathrm{RF} \mathbf{f}_{\mathrm{BB},n}-\zeta_{m,n}\right |^{2} \label{eqn:system optimization_1_1}\tag{10a}\\
&\qquad \qquad \,\, {\text {s.~t. }} \qquad \,\text{(\ref{constr:SINR_1}), (\ref{constr:HBF}), (\ref{constr:TP_1})}, \tag{10b}
\end{align}
where $\alpha>0$ is the penalty factor, which balances the objective function (\ref{eqn:system optimization}) and the equality constraint (\ref{constr:SINR_2}).
\subsection{Optimization of $\mathbf{F}_\mathrm{RF}$ for a given $\mathbf{F}_\mathrm{BB}$ and $\zeta_{m,n}$}
Let us define $\mathbf{w}\in \mathbb{C}^{{M_\mathrm{t}}N_\mathrm{t}\times1}$ and $\mathbf{X}_m\in \mathbb{C}^{N_\mathrm{t}\times {M_\mathrm{t}}N_\mathrm{t}}$ as
\begin{align}
 \mathbf{w}\overset{\Delta}{=}&[\mathbf{f}_{\mathrm{RF},1}^T,\mathbf{f}_{\mathrm{RF},2}^T,…,\mathbf{f}_{\mathrm{RF},{M_\mathrm{t}}}^T ]^T, \tag{11}\\
 \mathbf{X}_m\overset{\Delta}{=}&[\mathbf{f}_{\mathrm{BB},m}(1)\mathbf{I}_{N_\mathrm{t}}, \hdots, \mathbf{f}_{\mathrm{BB},m}({M_\mathrm{t}})\mathbf{I}_{N_\mathrm{t}}], \tag{12}
 \end{align}
 where $|\mathbf{w}(j)|=1, \forall j$. Thereby, the quantity $\mathbf{F}_\mathrm{RF}\mathbf{f}_{\mathrm{BB},m}$ can be rewritten as 
 \begin{align}
     \mathbf{F}_\mathrm{RF}\mathbf{f}_{\mathrm{BB},m}=\mathbf{X}_m\mathbf{w}, \tag{13}
     \end{align} 
which reduces the OP (10), for fixed values of $\mathbf{F}_\mathrm{BB}$ and $\zeta_{m,n}$, to the following equivalent formulation that depends only on $\mathbf{w}$ as
\begin{align}
&\min \limits_{\mathbf{w}} \quad f(\mathbf{w}) = - \sum_{l=1}^L\sum_{m=1}^{M_\mathrm{t}}w_l\left\vert\mathbf{a}^H_\mathrm{t}(\theta_l) \mathbf{X}_m\mathbf{w}\right\vert^2 \notag\\
&\qquad\qquad\qquad + \frac{\alpha}{2}\sum_{n=1}^{M_\mathrm{t}} \sum_{m=1}^{M_\mathrm{t}}\left |{\mathbf{h}^H_m \mathbf{X}_n \mathbf{w}-\zeta_{m, n}}\right |^{2} \tag{14a}\\
&{\text {s.~t.}} \quad {|\mathbf{w}(j)|=1,\quad \forall j }. \tag{14b}
\end{align}
The above OP is still non-convex due to the constant modulus constraint on the elements of $\mathbf{w}$. To solve this problem, we employ the highly efficient Riemannian conjugate gradient (RCG) algorithm \cite{mm_2}, which takes advantage of the Riemannian gradient to evaluate the descent direction. To this end, the Euclidean gradient of the function $f(\mathbf{w})$ is formulated as
\begin{equation}\label{phi_8}
\begin{aligned}
\nabla f(\mathbf{w}) = &2\sum_{n=1}^{M_\mathrm{t}} \sum_{m=1}^{M_\mathrm{t}}\mathbf{X}^H_n\mathbf{h}_m\left(\mathbf{h}^H_m\mathbf{X}_n\mathbf{w}-\zeta_{m,n}\right) \\
&-\alpha\sum_{l=1}^L\sum_{m=1}^{M_\mathrm{t}} w_l\mathbf{X}^H_m\mathbf{a}_\mathrm{t}(\theta_l)\mathbf{a}^H_\mathrm{t}(\theta_l) \mathbf{X}_m\mathbf{w}.
\end{aligned}
\tag{15}
\end{equation}   
Thus, we obtain the Riemannian gradient at the point $\mathbf{w}(j)$, which is defined as the orthogonal projection of $\nabla f(\mathbf{w})$ onto the tangent space of the manifold at the associated point $\mathbf{w}(j)$, yielding
\begin{equation*} 
\mathrm {grad}~ f\left[\mathbf{w}\left(j\right)\right]=\nabla f\left[\mathbf{w}\left(j\right)\right] -\Re\left\{{\nabla f\left[\mathbf{w}(j)\right] \odot \mathbf{w}(j)^{*}}\right \} \odot \mathbf{w}(j).
\tag{16}
\end{equation*}
Subsequently, the update rule of the search direction in the manifold space is given by 
\begin{equation} \label{RCG_5}
\begin{aligned}
\boldsymbol{\eta }^{i+1}=-\mathrm {grad}~ f(\mathbf{w}^{i+1}(j))+\nu \mathcal {T}_{\mathbf{w}^i(j) \rightarrow \mathbf{w}^{i+1}(j)}\left ({\boldsymbol {\eta }^{i}}\right),
\end{aligned}
\tag{17}
\end{equation}
where $\boldsymbol {\eta}^{i}$ denotes the search direction at $\mathbf{w}^{i}(j)$ in the $i$th iteration, $\nu$ is the update parameter chosen as the Polak-Ribiere parameter, and $\mathcal {T}_{\mathbf{w}^i(j) \rightarrow \mathbf{w}^{i+1}(j)}\left ({\boldsymbol {\eta }^{i}}\right)$ represents the operation mapping the search direction from its original
tangent space to the current tangent space. Finally, we carry out the retraction operation to determine the destination on the manifold.
\subsection{Optimization of $\mathbf{F}_\mathrm{BB}$ for a given $\mathbf{F}_\mathrm{RF}$ and $\zeta_{m,n}$} 
For fixed values of $\mathbf{F}_\mathrm{RF}$ and $\zeta_{m,n}$, the OP constructed for finding $\mathbf{F}_\mathrm{BB}$ is given by
\begin{align}
&\min \limits_{{\mathbf{F}_\mathrm{BB}}}\Upsilon\left(\mathbf{F}_\mathrm{BB}\right) = -\sum_{l=1}^L\sum_{m=1}^{M_\mathrm{t}}w_l\left\vert\mathbf{a}^H_\mathrm{t}(\theta_l) \mathbf{F}_\mathrm{RF}\mathbf{f}_{\mathrm{BB},m}\right\vert^2 \notag\\
&\qquad \qquad \qquad + \frac {\alpha}{2} \sum_{n=1}^{M_\mathrm{t}} \sum_{m=1}^{M_\mathrm{t}} \left | \mathbf{h}_{m}^{H} \mathbf{F}_\mathrm{RF} \mathbf{f}_{\mathrm{BB},n}-\zeta_{m,n}\right |^{2} \label{eqn:system optimization_1_1}\tag{18a}\\
&\qquad \qquad \,\, {\text {s.t. }} \qquad \,\text{(\ref{constr:TP_1})}. \tag{18b}
\end{align}
By ignoring the transmit power constraint (\ref{constr:TP_1}), $(18)$ reduces to an unconstrained OP. Thus, the optimal BB TPC $\mathbf{F}_\mathrm{BB}$ can be evaluated using the first-order optimality condition as follows
\begin{align}
\nabla \Upsilon \left(\mathbf{F}_\mathrm{BB}\right)&=0 \label{eqn:beta}\tag{19a}\\
\mathbf{f}_{\mathrm{BB},m}&=\alpha \mathbf{Y}^{-1} \sum_{n=1}^{M_\mathrm{t}} \tilde {\mathbf {h}}_{n} \zeta_{n,m},\quad \forall m , \tag{19b}  
\end{align}
where $\Tilde{\mathbf{h}}_n = \mathbf{F}^H_\mathrm{RF}\mathbf{h}_n \in \mathbb{C}^{M_\mathrm{t}\times 1}$ and $\mathbf{Y} =  \alpha \sum \limits_{n=1}^{M_\mathrm{t}}\Tilde{\mathbf{h}}_n\Tilde{\mathbf{h}}^H_n-2\mathbf{I}_{M_\mathrm{t}}$.

\subsection{Optimization of $\zeta_{m,n}$ for a given $\mathbf{F}_\mathrm{RF}$ and $\mathbf{F}_\mathrm{BB}$}
With the BB and RF TPCs fixed, the OP for $\zeta_{m,n}$ is given by
\begin{align}
&\hspace {-0.3pc}{\min \limits_{\zeta_{m,n}}} \quad { \sum_{n=1}^{M_\mathrm{t}} \sum_{m=1}^{M_\mathrm{t}} \left |{ \mathbf{h}_{m}^{H}\mathbf{F}_\mathrm{RF} \mathbf{f}_{\mathrm{BB},n}-\zeta_{m,n} }\right |^{2}} \tag{20a}\\
&\,{\text {s.t. }} \quad {\frac { \left |{ \zeta_{m,m} }\right |^{2}} {\sum_{n \neq m}^{K}\left |{\zeta_{m,n} }\right|^{2}+\sigma_{m}^{2}} \geq \Gamma_{m},\quad \forall m \in \mathcal {M}}. \label{constr:SOCP}\tag{20b}
\end{align}
Although the objective function in (20) is convex with respect to $\zeta_{m, n}$, the constraints are non-convex. To solve this problem, we employ the SOCP algorithm \cite{mm_ISAC_1}, where the constraints in (\ref{constr:SOCP}) can be expressed as second-order cones given by
\begin{align} 
\sqrt {1+\frac {1}{\Gamma_{m}}}\zeta_{m,n} \geq \left \|{\begin{array}{c} \mathbf{Z}^{H} \mathbf{c}_m \\ 
\sigma_{m} \end{array}}\right \|_{2},\quad \forall m ,\tag{21}
\end{align}
where $\mathbf{Z} \in \mathbb{C}^{M_\mathrm{t}\times M_\mathrm{t}}$ denotes a matrix with $\mathbf{Z}(m,n)=\zeta_{m,n}$ and $\mathbf{c}_m \in \mathbb{C}^{M_\mathrm{t} \times 1}$ is a vector with one as its $m$th element and other entries as zeros. Furthermore, the optimal weighing factor $w_l$ is evaluated using the minimum estimation mean square error (MSE) criterion \cite{weigh}, which is formulated as
\begin{align*}\label{eqn:weigh}
&w_l =
\begin{cases}
\frac {\dfrac {\chi^2(\theta_L)}{\chi^2(\theta_l)}}{{1 + \sum ^{L - 1}_{i = 1} \dfrac {\chi^2(\theta_L)}{\chi^2(\theta_i)}} }\quad ~1 \leq l \leq L - 1,\\
1 - \sum ^{L - 1}_{i = 1} w_i, \qquad \qquad ~~i = L,
\end{cases}
\tag{22}
\end{align*}
where $\chi(\theta_l)=\sum_{m=1}^{M_\mathrm{t}}\left\vert\mathbf{a}^H_\mathrm{t}(\theta_l) \mathbf{F}_\mathrm{RF}\mathbf{f}_{\mathrm{BB},m}\right\vert^2$. We summarize the key steps of the proposed method in Algorithm \ref{alg:algo_1}. Furthermore, the computational complexity of Algorithm \ref{alg:algo_1} is given by $\mathcal{O}\bigg(I_\mathrm{out}I_\mathrm{in}\Big(M^3_\mathrm{t}+I_\mathrm{r}\left(LM^3_\mathrm{t}N_\mathrm{t}+M^4_\mathrm{t}N_\mathrm{t}\right)+M^{3.5}_\mathrm{t}\Big)\bigg)$, where $I_\mathrm{r}$ is the number of iterations of the RCG algorithm required for updating $F_\mathrm{RF}$, and $I_\mathrm{out}$ and $I_\mathrm{in}$ denote the number of outer and inner iteration required for convergence, respectively. 
\begin{algorithm}[t]
\caption{Penalty-based triple alternating optimization algorithm to solve (7)}
\label{alg:algo_1}
\begin{algorithmic}[1]
    \State Initialize $\alpha, \mathbf{F}_\mathrm{RF}$, $\zeta_{m,n}, \forall m,n $ and stopping parameter $\epsilon>0$  
    \Repeat
            \Repeat
            \State Update $\mathbf{F}_\mathrm{RF}$ by solving the OP $(14)$
            \State Update $\mathbf{F}_\mathrm{BB}$ by $(20b)$
            \State Update $\zeta_{m,n}$ by solving the OP $(20)$
            \Until The objective function $\Psi\Big(w_l, \mathbf{F}_\mathrm{RF}, \mathbf{F}_\mathrm{BB}\Big)$ saturates
    \State Update $\alpha=\frac{\alpha}{e}, 0<e<1$.
    \Until $\left | \mathbf{h}_{m}^{H} \mathbf{F}_\mathrm{RF} \mathbf{f}_{\mathrm{BB},n}-\zeta_{m,n}\right |^{2}<\epsilon, \forall m, n$
    \State Update $\mathbf{F}_\mathrm{BB} = \sqrt{P_\mathrm{t}}\frac{\mathbf{F}_\mathrm{BB}}{\|\mathbf{F}_\mathrm{RF}\mathbf{F}_\mathrm{BB}\|_F}$
     \State Evaluate $w_l$ using (\ref{eqn:weigh})
    \State {\bf return} $\mathbf{F}_\mathrm{RF}, \mathbf{F}_\mathrm{BB}$
\end{algorithmic}
\end{algorithm}
\section{\uppercase{Simulation Results}}\label{simulation results}
To evaluate the performance of the proposed scheme for the MBOL model, we consider an ISAC BS equipped with $N_\mathrm{t}=64$ and $M_\mathrm{t}=6$ RF chains that serve an unobstructed two-way urban road. There are three CUs, which are located at angles of $-60^\circ, -40^\circ$ and $ -20^\circ$. Furthermore, the pathloss model $PL(d_m)$ of the mmWave MIMO channel is given as \cite{mm_2}
\begin{equation}\label{eqn:path loss model}
\begin{aligned}
PL(d_m)\hspace{0.02in}[{\rm dB}] = \varepsilon + 10\varphi\log_{10}(d_m)+\varpi,
\end{aligned}
\tag{23}
\end{equation}
where $\varpi \in {\cal CN}(0,\sigma_{\rm \varpi}^2)$ with $\sigma_{\rm \varpi}=5.8 \hspace{0.02 in}{\rm dB}$, $\varepsilon=61.4$ and $\varphi=2$ \cite{mm_2}.  
Additionally, we fix the number of multi-path components to $N_m^{\rm p}=10, \forall m$. Moreover, the angles of departure $\theta_{m,l}, \forall m, l$ are generated from a truncated Laplacian distribution with a uniformly-random mean angle of $\overline{\theta}$ and a constant angular spread of $\frac{\pi}{2N_\mathrm{t}}$. We assume the object to be located within a range of $20$m from the ISAC BS and the desired beam pattern to span the angular range $\left[30^\circ, 50^\circ\right]$. The reflection coefficient and the noise variance are set to $\tau_l=1, \forall l$ and $\sigma^2=-91\hspace{0.02 in}{\rm dBm}$, respectively. The penalty factor is initialized as $\alpha = 10^{-3}$.
We averaged the results over 500 independent channel realizations and compared our proposed scheme to three benchmarks: radar-only, ISAC-AB \cite{mm_ISAC_8}, and two-stage design \cite{mm_ISAC_1}.
\begin{figure}[t]
\centering
\includegraphics[width = 8 cm]{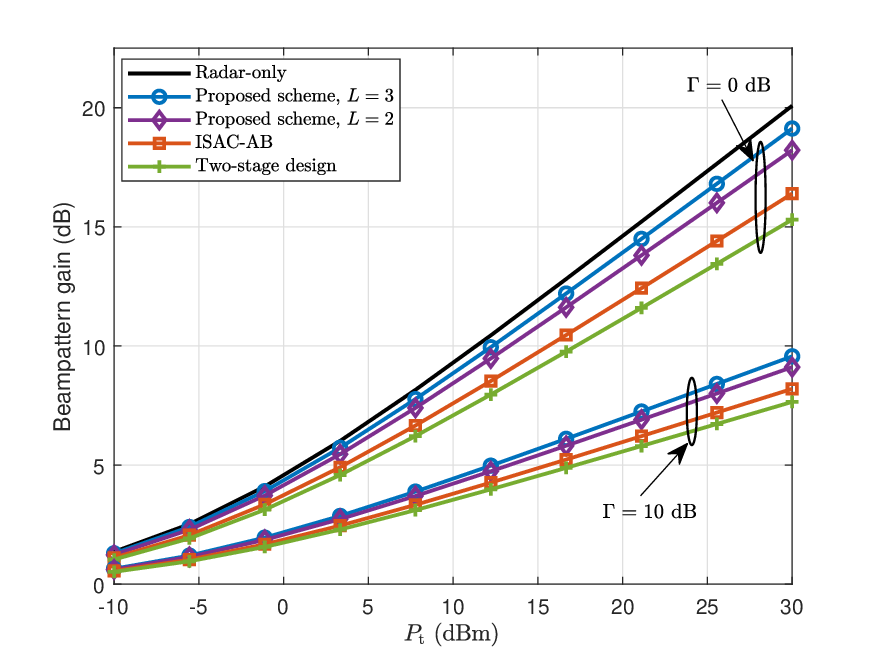}
\caption{SBP gain versus transmit power.}
\label{fig:R1}
\end{figure}

\begin{figure}[t]
\centering
\includegraphics[width = 8 cm]{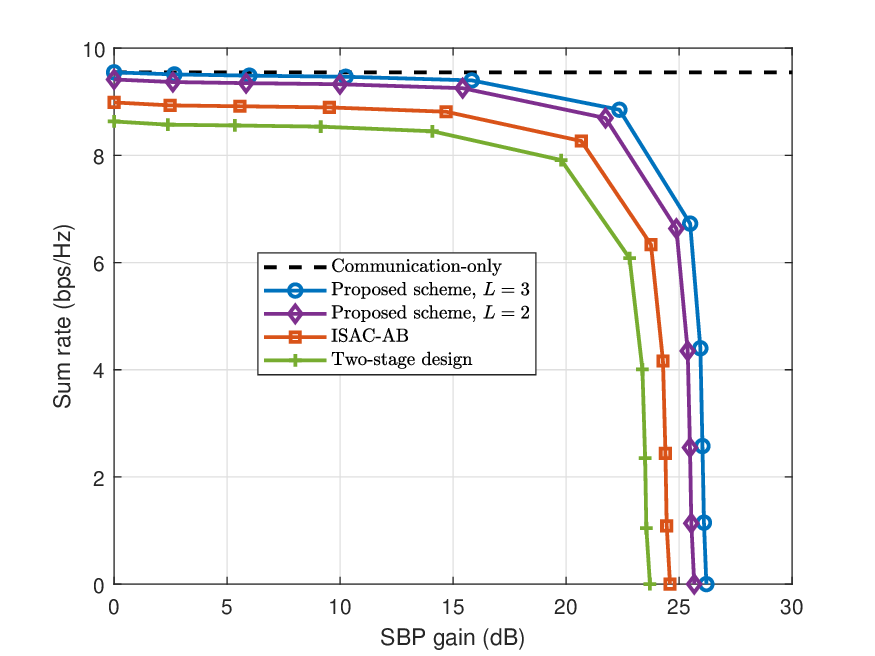}
\caption{Sum rate versus SBP gain.}
\label{fig:R2}
\end{figure}




In Fig. \ref{fig:R1}, we plot the SBP gain vs. transmit power $P_\mathrm{t}$ for different SINR thresholds $\Gamma \in \{0, 10\}$dB. As seen from the figure, the SBP gain of the system increases with $P_\mathrm{t}$ due to an improvement in the resultant gain at the object for higher values of $P_\mathrm{t}$. 
Furthermore, the proposed scheme outperforms both the ISAC-AB and the two-stage benchmarks for both $L=2$ and $L=3$. This arises due to the fact that both the ISAC-AB and two-stage methods require a higher level of power to generate the wider but lower-gain beam required for a closely located object, which reduces the array gain. In contrast, the proposed scheme enjoys a higher array gain resulting from the multiple narrow beams. 

Fig. \ref{fig:R2} depicts the trade-off between sensing and communication by plotting the sum rate versus the SBP gain of the system. We calculate the sum rate of the CUs as $\sum_{m=1}^M\log_2{\left(1+\mathrm{SINR}_m\right)}$, where transmit power is set to $10$ dBm. For comparison, we consider a communication-only scheme as a benchmark scheme, where the available power is used only to serve the CUs. As seen from the figure, the proposed scheme with $L=2$ and $3$ provides a better trade-off than the benchmark schemes. This is because multiple sharp beams in the proposed MBOL generate a higher array gain, and hence, more power is available to serve CUs. 

\begin{figure}[t]
\centering
\includegraphics[width = 8 cm]{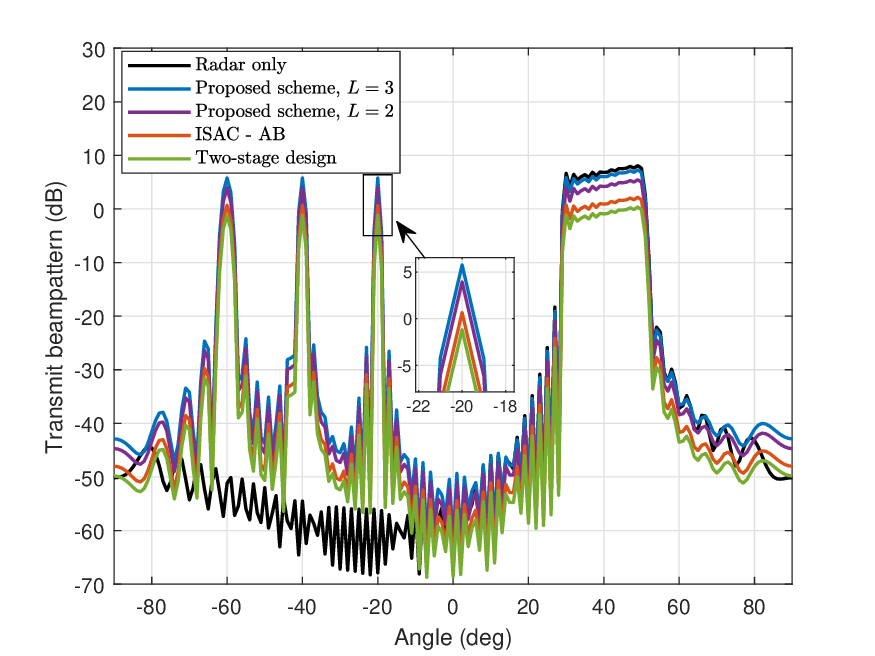}
\caption{Transmit beampattern with $P_\mathrm{t}=10$ dBm and $ \Gamma=0$ dB.}
\label{fig:R3}
\end{figure}

Furthermore, to characterize the tradeoff achieved by the proposed design between the radar
and communication systems, we plot the transmit beampattern in Fig. \ref{fig:R3}. As seen from the figure, there are three peaks at the CUs in addition to a wide lobe at the object. As $L$ increases, the transmit beampattern of the proposed scheme approaches that of the ideal radar-only benchmark. 

\begin{figure}[t]
\centering
\includegraphics[width = 8 cm]{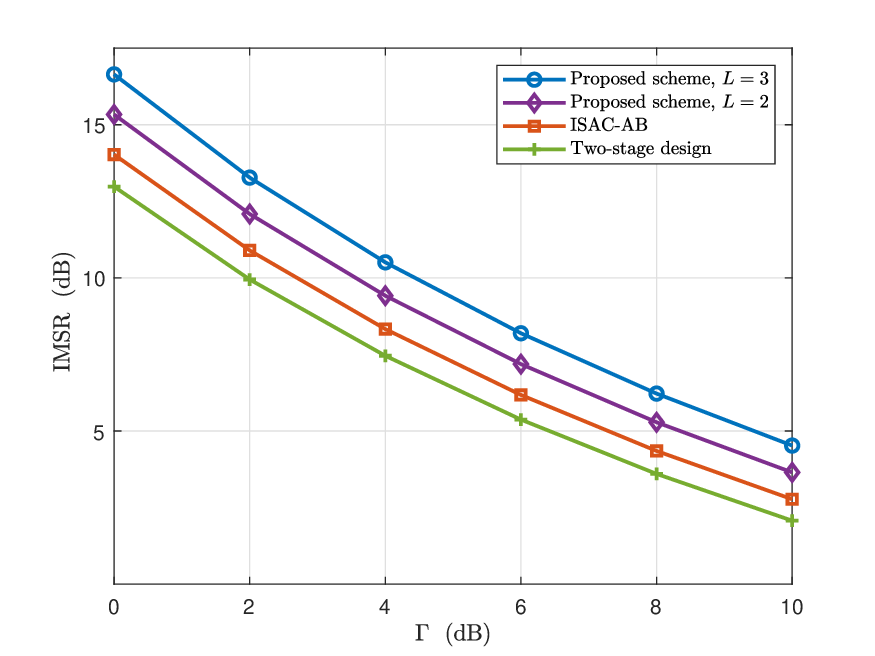}
\caption{IMSR versus SINR threshold with $P_\mathrm{t}=10$ dBm.}
\label{fig:R4}
\end{figure}

To include the impact of sidelobe gains in the sensing performance, we further investigate the integrated-mainlobe-to-sidelobe-ratio (IMSR) in Fig. \ref{fig:R4}. We calculate the IMSR according to \cite{rev_3}, where mainlobe and sidelobe regions are $\big[\theta_0\pm \frac{\Delta}{2}\big]$ and $\big[-\frac{\pi}{2},\theta_0-\frac{\Delta}{2}\big]\cup\big[\theta_0+\frac{\Delta}{2},\frac{\pi}{2}\big]$ with $\theta_0 = 40^\circ$ and $\Delta=20^\circ$. Observe that the IMSR decreases upon increasing $\Gamma$, because less power is radiated towards the object for guaranteeing the SINR requirements of the users. Moreover, the IMSR of the proposed scheme is better than that of the benchmark schemes as a benefit of having multiple energy-efficient pencil sharp beams, which makes the MBOL model ideally suited for CAV applications.

Note that the proposed MBOL is remarkably suitable for vehicular applications of ISAC systems, when there is a rapid variation of the angle of departure (AoD). Consequently, one has to form an ultra-wide beam in the conventional methods to track the object, which leads to a loss of a significant amount of energy owing to the reduced beamforming gain. 
By contrast, the energy-efficient pencil-sharp beams of the proposed MBOL model can accurately track high-mobile objects by increasing the number of beams.
\section{\uppercase{Conclusion}}\label{conclusion}
An SBP gain optimization problem was formulated for a mmWave MIMO ISAC-aided CAV system based on the MBOL model. A penalty-based triple alternating optimization scheme was proposed in which the variables are decoupled using the penalization method. Subsequently, the alternating optimization paradigm, which involves the RCG and SOCP procedures, was adopted to design the RF and BB TPCs. The simulation results showed the effectiveness of the proposed MBOL model and optimization scheme in ultra-narrow pencil beam mmWave systems to enhance the SBP gain of closely located objects.
\newpage
\bibliographystyle{IEEEtran}
\bibliography{biblio.bib}

\begin{thebibliography}{10}
\providecommand{\url}[1]{#1}
\csname url@samestyle\endcsname
\providecommand{\newblock}{\relax}
\providecommand{\bibinfo}[2]{#2}
\providecommand{\BIBentrySTDinterwordspacing}{\spaceskip=0pt\relax}
\providecommand{\BIBentryALTinterwordstretchfactor}{4}
\providecommand{\BIBentryALTinterwordspacing}{\spaceskip=\fontdimen2\font plus
\BIBentryALTinterwordstretchfactor\fontdimen3\font minus
  \fontdimen4\font\relax}
\providecommand{\BIBforeignlanguage}[2]{{%
\expandafter\ifx\csname l@#1\endcsname\relax
\typeout{** WARNING: IEEEtran.bst: No hyphenation pattern has been}%
\typeout{** loaded for the language `#1'. Using the pattern for}%
\typeout{** the default language instead.}%
\else
\language=\csname l@#1\endcsname
\fi
#2}}
\providecommand{\BIBdecl}{\relax}
\BIBdecl

\bibitem{rev_3}
Z.~Cheng, C.~Han, B.~Liao, Z.~He, and J.~Li, ``Communication-aware waveform
  design for {MIMO} radar with good transmit beampattern,'' \emph{IEEE Trans.
  Signal Process.}, vol.~66, no.~21, pp. 5549--5562, 2018.

\bibitem{mm_ISAC_4}
X.~Yu, L.~Tu, Q.~Yang, M.~Yu, Z.~Xiao, and Y.~Zhu, ``Hybrid beamforming in
  {mmWave} massive {MIMO} for {IoV} with dual-functional radar communication,''
  \emph{” IEEE Trans. Veh. Technol.}, vol.~72, no.~7, pp. 9017--9030, 2023.

\bibitem{cha_est}
A.~Gupta, M.~Jafri, S.~Srivastava, A.~K. Jagannatham, and L.~Hanzo, ``An affine
  precoded superimposed pilot-based {mmWave MIMO-OFDM ISAC} system,''
  \emph{IEEE Open Journal of the Commun. Society}, vol.~5, pp. 1504--1524,
  2024.

\bibitem{mm_2}
J.~Singh, S.~Srivastava, S.~P. Yadav, A.~K. Jagannatham, and L.~Hanzo, ``Joint
  hybrid transceiver and reflection matrix design for {RIS}-aided {mmWave MIMO}
  cognitive radio systems,'' \emph{IEEE Transactions Cognitive Commun. and
  Networking}, pp. 1--1, 2024.

\bibitem{mm_ISAC_5}
F.~Liu, W.~Yuan, C.~Masouros, and J.~Yuan, ``Radar-assisted predictive
  beamforming for vehicular links: Communication served by sensing,''
  \emph{IEEE Trans. Wireless Commun.}, vol.~19, no.~11, pp. 7704--7719, 2020.

\bibitem{mm_ISAC_6}
W.~Yuan, F.~Liu, C.~Masouros, J.~Yuan, D.~W.~K. Ng, and N.~González-Prelcic,
  ``Bayesian predictive beamforming for vehicular networks: A low-overhead
  joint radar-communication approach,'' \emph{IEEE Trans. Wireless Commun.},
  vol.~20, no.~3, pp. 1442--1456, 2021.

\bibitem{mm_ISAC_7}
X.~Meng, F.~Liu, C.~Masouros, W.~Yuan, Q.~Zhang, and Z.~Feng, ``Vehicular
  connectivity on complex trajectories: Roadway-geometry aware {ISAC}
  beam-tracking,'' \emph{IEEE Trans. Wireless Commun.}, pp. 1--1, 2023.

\bibitem{mm_ISAC_8}
Z.~Du, F.~Liu, W.~Yuan, C.~Masouros, Z.~Zhang, S.~Xia, and G.~Caire,
  ``Integrated sensing and communications for {V2I} networks: Dynamic
  predictive beamforming for extended vehicle targets,'' \emph{IEEE Trans.
  Wireless Commun.}, vol.~22, no.~6, pp. 3612--3627, 2023.

\bibitem{mm_ISAC_1}
C.~Qi, W.~Ci, J.~Zhang, and X.~You, ``Hybrid beamforming for millimeter wave
  {MIMO} integrated sensing and communications,'' \emph{IEEE Commun. Letters},
  vol.~26, no.~5, pp. 1136--1140, 2022.

\bibitem{rev_1}
Z.~Du, F.~Liu, Y.~Li, W.~Yuan, Y.~Cui, Z.~Zhang, C.~Masouros, and B.~Ai,
  ``Towards {ISAC}-empowered vehicular networks: Framework, advances, and
  opportunities,'' 2023.

\bibitem{ISAC_3}
H.~Hua, J.~Xu, and T.~X. Han, ``Optimal transmit beamforming for integrated
  sensing and communication,'' \emph{” IEEE Trans. Veh. Technol.}, vol.~72,
  no.~8, pp. 10\,588--10\,603, 2023.

\bibitem{weigh}
C.~Xing, Y.~Jing, and Y.~Zhou, ``On weighted {MSE} model for {MIMO} transceiver
  optimization,'' \emph{IEEE Trans. Veh. Technol.}, vol.~66, no.~8, pp.
  7072--7085, 2017.

\end{thebibliography}
\end{document}